\documentstyle[twocolumn,aps,prl,graphics]{revtex}

\begin{document}

\twocolumn[\hsize\textwidth\columnwidth\hsize\csname
@twocolumnfalse\endcsname

\title{\Large {\em Ab initio} evaluation of the charge-ordering
in  $\alpha^\prime N\!aV_2O_5$ }

\author{Nicolas Suaud and Marie-Bernadette Lepetit}

\address
{Laboratoire de Chimie Quantique et Physique Mol\'eculaire, \\
Unit\'e Mixte de Recherche 5626 du CNRS,\\
118 route de  Narbonne, F-31062 Toulouse, France}

\date{\today}

\maketitle

\begin{abstract}
We report {\it ab initio} calculations of the charge ordering in
$\alpha^\prime N\!aV_2O_5$ using large configurations interaction
methods on embedded fragments. Our major result is that the $2p_y$
electrons of the bridging oxygen of the rungs present a very strong
magnetic character and should thus be explicitly considered in any
relevant effective model. The most striking consequence of this result
is that the spin and charge ordering differ substantially, as differ
the experimental results depending on whether they are sensitive to
the spin or charge density.

\smallskip
\end{abstract}

\vskip2pc]

The problem of charge ordering in the $\alpha^\prime N\!aV_2O_5$
compound has been widely discussed since Isobe {\it et
al}~\cite{Isobe} discovered in 1996 that this compound presents a
spin-Peierls transition at $34K$. From the first X-ray structural
determination by Carpy and Galy~\cite{Galy} a charge ordering was
assumed in the compound. Indeed, the $P2_1mn$ non-centrosymmetric group
yielding two inequivalent vanadium atoms, it supported the hypothesis of
an electronic structure formed by an alternation of magnetic $V^{4+}$
chains and non-magnetic $V^{5+}$ chains. When a spin-Peierls
transition was found its interpretation through the
dimerization of the magnetic chains seemed quite obvious.  In 1998,
the reinvestigation of the high temperature (HT) crystal structure~\cite{Pmmn}
yielded a $Pmmn$ centrosymmetric group for which the equivalence of
all vanadium atoms lifted any grounding to the charge ordering. 
The controversy was raised again by Damascelli {\it et al.} in
1998~\cite{Damascelli}. The strong on-rung doublet-doublet absorption
band, observed in the optical conductivity spectrum around $\Delta E =
1eV$, was attributed by the authors to a charge transfer excitation of
the magnetic electron between the two vanadium atoms. This picture
assumes a strong charge order on the rung (of about $0.8$
electron), however both the crystallographic data  and the $^{51}V$ NMR
studies~\cite{Ohama1} exhibit only one type of vanadium atom.

It is now widely accepted that the $\alpha^\prime N\!aV_2O_5$
spin-Peierls transition is in fact a double transition~\cite{Koppen},
consisting in a magneto-distortion of the lattice following a charge
ordering of the unpaired electrons.  X-ray~\cite{Ludecke} and electron
diffraction~\cite{Tsuda} experiments leaded to propose for the low
temperature (LT) phase a $Fmm2$ symmetry group associated with a doubling
of the unit cell in the $a$ and $b$ directions while it is quadrupled
in the $c$ direction. This low temperature symmetry yields two
inequivalent, but very similar, $V_2O_5$ layers, each of them
containing three inequivalent vanadium atoms.  $^{51}V$ NMR
experiments~\cite{Ohama1} see however only two inequivalent sets of
vanadium atoms.  In order to explain these experimental results, a
large number of hypotheses have been proposed for the electronic
structure of the low temperature phase. Among these
hypotheses the partisans of a total charge ordering ($V^{5+}$ and
$V^{4+}$ sites) are the most numerous. They suppose either the same
alternation of magnetic $V^{4+}$ and diamagnetic $V^{5+}$ chains that
was first suggested for the high temperature phase, or a zig-zag
charge ordering along the ladders~\cite{zigzag}. More in agreement
with the X-ray structural symmetries, but not with the NMR results, is
the hypothesis of two alternated ladders, one supporting two
equivalent $V^{4.5+}$ atoms and the other a zig-zag $V^{5+}$--$V^{4+}$
charge ordering. Fewer authors suggest that the charge ordering is
only partial~\cite{Fagot} and that one should consider the previous
charge localizations hypotheses with a $V^{4.5\pm \delta}, \; \delta
\ll 0.5$ charge ordering.

The aim of the present {\em ab-initio} quantum chemical calculations
is to investigate the local electronic structure of the compound in
order to better understand ---~and solve some of the
controversies of~--- the charge ordering both in high and low
temperature phases.

For this purpose we used a method of {\em ab-initio} spectroscopy of
{\em embedded} crystal fragments, using large atomic basis
sets~\cite{base} and large differential configurations interactions
methods dedicated to the accurate evaluation of both transitions
energies and states wave-functions~\cite{IDDCI} (for a detailed
description of the method applied to this system see our paper on the
evaluation of $t-J$ and two-band Hubbard models for the high
temperature phase~\cite{Nous}). This method has been successfully used
both in molecular magnetic systems~\cite{molmagn} and in strongly
correlated materials~\cite{CX} in order to determine super-exchange or
hopping integrals with an exceptional accuracy.  The bath is designed
in order to model the major effects (Madelung and Pauli exclusion) of
the rest of the crystal on the fragment, that is in such a way that
the crystal density of states projected on the atomic orbitals (the
most crucial being the magnetic and bridging orbitals) is accurately
reproduced in the embedded fragment. This goal is reached
through a combination of punctual charges and total-ions
pseudo-potentials (TIP) (see ref.~\cite{Nous} for more details and
validity checks on the high temperature $\alpha^\prime N\!a V_2O_5$
phase).

We have computed all (fundamental-)doublet to (first excited-)doublet
excitation energies and associated wave-functions, on all
non-equivalent pairs of nearest neighbors vanadium atoms, in the high
and low temperature phases. The crystal fragments are composed of two
adjacent $VO_5$ pyramids (corner sharing for the rungs and along the
ladders, edge-sharing between the ladders).  The positions of the
atoms as well as the bath charges and TIP are taken from the X-ray
crystallographic data~\cite{Pmmm,Ludecke}.

The most important results of our calculations, both in the high and
low temperature phases is that \begin{itemize}
\item the magnetic sites are not the vanadium
atoms by themselves but the whole $V-O-V$ rungs (as first suggested by
Horsch~\cite{Horsch}), 
\item these magnetic sites do not support one  delocalized 
or partly delocalized magnetic $3d$ electron but that the $2p$
electrons of the bridging oxygen atoms ---~supported by the $2p_y$
orbital pointing along the $b$ axis~--- present also a strong magnetic
character.
\end{itemize}
Indeed the computed wave-functions (reported below for the HT and LT
fundamental doublets) for the fundamental, as well as first excited,
doublet states exhibit a strong contribution on the configurations
where the $2p_y$ orbital of the bridging oxygen is half-filled.  In
fact the magnetic problem on the rungs should not be considered as one
magnetic electron delocalized between the $3d_{xy}$ of the two
vanadium atoms but rather as a resonance between this configuration
($\alpha |d_ip\bar{p}\rangle + \beta |p\bar{p}d_j\rangle$ where $d_i$
and $d_j$ respectively stand for the magnetic $3d_{xy}$ orbitals of
the two vanadium atoms and $p$ for the $2p_y$ orbital of the bridging
oxygen atom) and the configuration with three magnetic electrons
respectively localized on the magnetic $3d_{xy}$ orbitals of two vanadiums
and the $2p_y$ orbital of the bridging oxygen ($-\left(\gamma+\delta
\right)|d_i\bar{p}d_j\rangle +\gamma |\bar{d_i}pd_j\rangle+ \delta
|d_ip\bar{d_j}\rangle $).

The wave function of the HT fundamental doublet of the rungs is
\begin{eqnarray*} \label{D+HT}
& &
+0.63\left(|d_ip\bar{p}\rangle +|p\bar{p}d_j\rangle\right)/\sqrt{2} \\
&& +0.68\left(-2|d_i\bar{p}d_j\rangle +| \bar{d_i}pd_j\rangle+
|d_ip\bar{d_j}\rangle \right)/\sqrt{6}  \\ 
&& +0.13\left(|d_i\bar{d_i}d_j\rangle+| d_id_j\bar{d_j}\rangle \right)/ \sqrt{2}  \\
&&+0.12\left(|d_i\bar{d_i}p\rangle-|pd_j\bar{d_j}\rangle \right)/ \sqrt{2}\\ 
&& + {\rm small\; terms\;} \ldots  
\end{eqnarray*}
The LT phase is known~\cite{Ludecke} for presenting an alternation
(along the $\vec a$ direction) between modulated and non-modulated
ladders as well as two slightly different $(a,b)$ planes. While one
out of two ladders remains of similar geometry as in the high
temperature phase, the other ladder takes a small zig-zag form along
the $\vec b$ direction. The results on the non-modulated rungs are very similar as in the HT phase and will not be further detailed. 
The wave functions of the LT fundamental doublets on the modulated rungs are
\begin{eqnarray*} \label{D+BTM}
&&
+0.25 (0.23)|d_ip\bar{p}\rangle + 0.59(0.60)|p\bar{p}d_j\rangle \\
&&+0.53(0.52)|d_i\bar{p}d_j\rangle -0.40(0.41)|\bar{d_i}pd_j\rangle-0.13(0.11)|d_ip\bar{d_j}\rangle \\ 
&& +0.11(0.11)|d_i\bar{d_i}d_j\rangle+0.06(0.06)| d_id_j\bar{d_j}\rangle \\
&&-0.06(0.05)|d_i\bar{d_i}p\rangle+0.09(0.09)|pd_j\bar{d_j}\rangle \\&& + 
{\rm small\; terms\;} \ldots  
\end{eqnarray*}
where the number in parentheses correspond to the alternate (a,b) plane. 

The first consequence of the strong magnetic character of the bridging
oxygen atom is that it should be explicitly considered while
modeling the compound. One of the most striking implication is that
while in the single magnetic electron picture the charge and spin
sites populations are equal when $3$ magnetic electrons are considered
the charge and spin sites populations can strongly differ.
Table~\ref{tbl:popmagn} displays the charge and spin population of the
magnetic orbitals.
\begin{table}[x]
\begin{tabular}{c|ddd|ddd}
&\multicolumn{3}{c|}{Charge populations}&\multicolumn{3}{c}{Spin populations} \\ \hline 
&$d_i$ & $p$ & $d_j$ &$d_i$ & $p$ & $d_j$ \\ \hline 
HT phase    & 0.785& 1.43& 0.785 & 0.58 & -0.16 & 0.58 \\
LT non Mod. & 0.78 & 1.44 & 0.78 & 0.575& -0.15 & 0.575 \\
LT Mod.     & 0.61 & 1.45 & 0.94 & 0.22 & -0.10 & 0.88 
\end{tabular}
\caption{Magnetic orbitals charge and spin populations on the rungs 
(in electrons units). Note that on the modulated rungs the $V_i$--$O$
distance is shorter than the $V_j$--$O$ distance.}
\label{tbl:popmagn}
\end{table}
One should notice that while the $3d$ charge population ordering on
the modulated rungs is only $2\delta_{ch}=0.33$, the participation of
the oxygen to magnetic properties results in a spin population
ordering nearly twice as large with $2\delta_{sp}=0.66$. On the
non-modulated rungs however, as expected from the crystallographic
data, there is not any charge ordering and the overall results are
very similar to the high temperature phase. One should however notice
that the charge and spin populations remain substantially different,
with a strong increase of the $3d$ population (compared to the $0.5$
value expected in the one magnetic electron picture) and a smaller increase
of the spin polarization of the $3d$ orbitals by inverse polarization
on the central oxygen $2p$ orbital. 

Another result worth to be pointed at is the substantial differences
we found between the magnetic orbitals charge populations and the
total atomic populations. It is well known that such total population
calculations are partly dependent to the choice in the atomic basis
set, and in particular to the spatial extension of the latter. It is
however widely accepted in quantum chemistry that the population
variations are significant and reliable evaluations of charge transfer
processes.  Table~\ref{tbl:mull} therefore presents the computed
atomic populations variations through the transition.
\begin{table}[h]
\begin{tabular}{c|ddd|ddd}
&\multicolumn{3}{c}{Non-modulated rungs} &\multicolumn{3}{c}{Modulated rungs} \\
&$V_i$ & $O$ & $V_j$& $V_i$ & $O$ & $V_j$ \\
\hline
Charge&-0.005 & +0.000 & -0.005 & +0.007 & -0.001 & -0.038 \\
Spin  &-0.002 & +0.005 & -0.002 & -0.37  & +0.05  & +0.32
\end{tabular}
\caption{Variation of the total atomic charge and spin populations 
(in number of
electrons units) in the low temperature phase with respect with the
high temperature phase. Note that on the modulated rungs the
$V_i$--$O$ distance is shorter than the $V_j$--$O$ distance.}
\label{tbl:mull}
\end{table}
One notices immediately that while the atomic total spin population
results are totally coherent with the magnetic orbital spin
populations reported in table~\ref{tbl:popmagn}, the atomic total
charge populations are substantially different from the magnetic
orbital charge populations. Indeed while the total charge
disproportion between the two vanadium atoms of the modulated rungs is
extremely small ($2\delta_{chtot}=0.045$), the disproportion between the
two $3d$ magnetic orbital populations is $2\delta_{ch}=0.33$. 
A finer analysis of the electrons repartition in term of atomic
orbitals shows that there are substantial electron population 
transfers between the vanadium magnetic $3d_{xy}$ orbitals and the
other $3d$ vanadium orbitals. 

Indeed, while the $V_j$ atom sees its total number of electron
decrease, the occupation of its $3d_{xy}$ atomic magnetic orbital
increases by $0.143$ electron due to a population transfer from the
$3d_{x^2-y^2}$, $3d_{z^2}$ and $3d_{xz}$ orbitals. On the $V_i$ atom
the population transfer goes in reverse order and the atomic magnetic
orbital looses $0.166$ electron in favor of the $3d_{x^2-y^2}$,
$3d_{z^2}$ and $3d_{xz}$ orbitals (see table~\ref{tbl:co1},~\ref{tbl:co2}).  
\begin{table}[h]
\begin{tabular}{ccccc}
$d_{xy}$ & $d_{x^2-y^2}$ & $d_{xz} $ & $d_{z^2}$ \\
\hline
0.309 & -0.136 & -0.088 & -0.085 
\end{tabular}
\caption{Orbital population difference between the $V_j$ and $V_i$
atoms of the modulated rungs in the low temperature phase. Units are
in number of electron.}
\label{tbl:co1}
\begin{tabular}{c|dddd}
Atom  & $d_{xy}$ & $d_{x^2-y^2}$ & $d_{xz} $ & $d_{z^2}$ \\
\hline
$V_i$ & -0.166   & +0.068        & +0.044    & +0.047 \\
$V_j$ & +0.143   & -0.068        & -0.044    & -0.038 
\end{tabular}
\caption{Difference between the $3d$ orbitals populations of the two
vanadiums of the rungs in the LT phase (modulated rungs) and the HT phase.}
\label{tbl:co2}
\end{table}
This electron transfer can be understood if one considers that the
$2p_x$ doubly-occupied orbital of the bridging $O$ atoms strongly
overlaps with the $3d_{x^2-y^2}$, $3d_{z^2}$ and $3d_{xz}$ orbitals of
the vanadiums. $V_i$ being closer to the oxygen than $V_j$ by
$0.13\AA$, the larger overlap between the oxygen $2p_x$ and the
vanadium $3d_{x^2-y^2}$, $3d_{z^2}$ and $3d_{xz}$ orbitals increases
the occupation of the later orbitals. The opposite phenomenum acts on
the $V_j$ atom.  Such an increase in the $3d$ atomic population would
result in a proportional increase of the on-site repulsion unless
another valence orbital, close to the Fermi level, is simultaneously
de-populated. This is precisely the case with the magnetic $3d_{xy}$
orbital. The combined result of this $d \rightarrow d$ transfers and
the magnetic character of the bridging oxygen results in a strong
difference between the total charge ($2\delta_{chtot}=0.05$) and the
spin ($2\delta_{sptot}=0.69$) orderings.

The consequence of this analysis, and this is one of the major
conclusions of the present work to draw this unexpected result, is
that experiments sensitive to the local charge density and experiments
sensitive to the local spin density should yield substantially
different results.

Summarizing our results on the charge ordering, we have found (in
agreement with the crystallographic data) that there is neither charge
nor spin ordering in the high temperature phase, that there is neither
charge nor spin ordering on the non-modulated rungs in the low
temperature phase and that while the total charge ordering is very
small on the modulated rungs, the spin ordering is on the contrary
quite large. 

In the HT phase, our results agree nicely both with the $^{51}V$ NMR
studies~\cite{Ohama1} (as them we see only one type of vanadium atom)
and optical conductivity measurements~\cite{Damascelli}. Indeed, our
computed on-rung doublet-doublet excitation energy is $1.08eV$ which
compares very well with the $1eV$ optical conductivity peak.  

In the LT phase Presura {\it et al.}~\cite{Presura} conclude from
optical conductivity experiments, that there should not be any major
charge redistribution through the phase transition, in full agreement
with our results. The apparent contradiction between these results and
the inelastic neutron scattering experiments~\cite{Grenier} that
predict large charge ordering can also be lifted. Indeed, the latter
are sensitive to the spin ordering which we have seen to be very
different from the charge ordering, due to the magnetic nature of the
bridging oxygen of the rung. Finally, the $^{51}V$ NMR
experiments~\cite{Ohama1} predict only two different types of vanadium
atoms in the LT phase. NMR experiments are sensitive to the electronic
spin polarization at the vanadium nuclei, that is essentially the spin
polarization of the vanadium $s$ orbitals induced by the spin
polarization of the magnetic sites. A simple energetic perturbative
argument shows that the major contribution to this $s$ spin
polarization must be supported by the vanadium $4s$ orbitals. Indeed,
the $4s$ orbitals are the only ones close enough to the Fermi level to
interact with the magnetic electrons in a non negligible way. We
therefore computed the induced spin polarization on the vanadiums $4s$
orbitals. On modulated rungs we found a spin $4s$ population of
$0.24\times 10^{-2}$ on $V_i$ and $1.51 \times 10^{-2}$ on $V_j$ while
on the non-modulated rungs we found $0.005\times 10^{-2}$, that is two
order of magnitude lower. As a title of comparison the spin vanadium
$4s$ polarization is in the HT phase $0.9\times 10^{-2}$. It seems
therefore possible to explain the NMR experiments in the low
temperature phase by the fact that the induced spin polarization on
the $s$ orbital of the non-modulated rungs vanadium atoms is so low
that their signal is not observed in the experimental window.

The main message we would like to send as a conclusion is that the
$\alpha^\prime N\!a V_2 O_5$ compound should not so much be considered
as a quarter-filled system with one magnetic electron delocalized on
the two vanadiums of each ladder rung, but rather as a resonating
system where the $V-O-V$ rungs support three magnetic electrons on
three magnetic orbitals, namely the two $d_xy$ orbitals of the
vanadium atoms and the $p_y$ orbital of the bridging oxygen. A major
consequence of this result is that, on the LT-modulated rungs, the
charge and spin ordering are different. Indeed, while the charge
ordering is very small, the spin ordering is quite large. The direct
implication is that experiments sensitive to the {\em spin} density
and experiments sensitive to the {\em charge} density {\em should} not
yield similar ordering amplitudes. We therefore think that a 
reinterpretation of experimental measurements in view of the present
results could solve most of their apparent inconsistencies.

\acknowledgments We thank Dr. D. Maynau for giving us graciously his
CASDI code, Dr. T. Ziman and Dr. V. Robert for fruitful discussions
and Prof. Van der Marel for his remarks.


\begin{thebibliography}{99}

\bibitem{Isobe}
M. Isobe and Y. Ueda, {\it J. Phys. Soc. Jnp.} {\bf 65}, 1178 (1996).

\bibitem{Galy} J. Galy, A. Casalot, M. Pouchard and P. Hagenmuller,
{\it C. R. Acad. Sc. Paris C} {\bf 262}, 1055 (1966);
A. Carpy and J. Galy, {\it Acta Cryst. B} {\bf 31}, 1481 (1975).

\bibitem{Pmmn} H. G. Von Schnering, Y. Grin, M. Kaupp, M. Somer
R. K. Kremer, O. Jepsen, T. Chatterji and M. Weiden, {\it
Z. Kristallogr.} {\bf 213}, 246 (1998); A. Meetsma, J. L. de Boer,
A. Damascelli, T. T. M. Palstra, J. Jegoudez and A. Revcolevschi, {\it
Acta. Cryst. C} {\bf 54}, 1558 (1998).


\bibitem{Damascelli} A. Damascelli, D. van der Marel, M. Gr\"uninger,
C. Presura, T. T. M. Palstra, J. Jegoudez and A. Revcolevschi,
{\it Phys. Rev. Lett.} {\bf 81}, 918 (1998).


\bibitem{Ohama1}T. Ohama, H. Yasuoka, M. Isobe and Y. Ueda,
{\it Phys. Rev. B} {\bf 59}, 3299 (1999).

\bibitem{Koppen} M. K\"oppen, D. Pankert, R. Hauptmann, M. Lang, M. Weiden,
C. Geibel, and F. Steglich,{\it Phys. Rev. B} {\bf 57}, 8466 (1998).

\bibitem{Ludecke} J. L\"udecke, A. Jobst, S. van Smaalen E. MorrŽ,
C. Geibel and H.-G. Krane, {\it Phys. Rev. Lett.} {\bf 82}, 3633
(1999);
A. Bernert, T. Chatterji, P. Thalmeier and P. Fulde, {\it Eur. Phys. J. B.}
{\bf 21}, 535 (2001). 

\bibitem{Tsuda} K. Tsuda, S. Amamiya, M. Tanaka, Y. Noda, M. Isobe and Y. Ueda,
{\it J. Phys. Soc Jpn.} {\bf 69}, 1935 (2000).

\bibitem{zigzag} P. Thalmeier and P. Fulde {\it Europhys. Lett.}
{\bf 44}, 242 (1998); T. Ohama, A. Goto, T. Shimizu, E. Ninomiya,
H. Sawa, M. Isobe and Y. Ueda,{\it J. Phys. Soc Jpn.} {\bf 69}, 2751
(2000).

\bibitem{Fagot} Y. Fagot-Revurat, M. Mehring and R. K. Kremer,
{\it Phys. Rev. Lett.} {\bf 84}, 4176 (2000).


\bibitem{base} The inner-core electrons ($[1s^22s^22p^63s^2]$ for the
$V$ atoms and $[1s^2]$ for the $O$) are represented by effective core
potentials (ECP). The outer-core and valence electrons are represented
using a $3s,3p,4d$ contracted basis set for the $V$ and a $2s4p1d$
basis set for the $O$ atoms. Exact expressions of the basis sets and
ECP can be found in ``Z. Barandiar\'an and L. Seijo, {\it Can. J. Chem.}
{\bf 70}, 409 (1992)''.



\bibitem{IDDCI} V. M. Garc\'ia, O. Castell, R. Caballol and J. P. Malrieu,
{\it Chem. Phys. Lett.} {\bf 238}, 222 (1995);
V. M. Garc\'ia, M. Reguero and R. Caballol, {\it
Theor. Chem. Acc.} {\bf 98}, 50 (1997); J. Miralles, J. P. Daudey and
R. Caballol, {\it Chem. Phys. Lett.} {\bf 198}, 555 (1992). 

\bibitem{Nous} N. Suaud and M. B. Lepetit, {\it Phys. Rev. B} {\bf
62}, 402 (2000).

\bibitem{molmagn} J. Cabrero, N. Ben Amor, C. de Graaf, F. Illas and
R. Caballol, {\it J. Phys. Chem.} {\bf A 104}, 9983 (2000).

\bibitem{CX} See for instance~: C. Jimenez Calzado, J. Fernandez Sanz,
J. P. Malrieu and F. Illas, {\it Chem. Phys. Lett.} {\bf 307}, 102
(1999);  D. Munoz, F. Illas, I. de P.R. Moreira, {\it Phys. Rev. Letters}
 {\bf 84}, 1579 (2000); C. Jimenez Calzado, J. Fernandez Sanz and
J. P. Malrieu, {\it J. Chem. Phys.} {\bf 112}, 5158 (2000).

\bibitem{Horsch} P. Horsch and F. Mack,
{\it Eur. Phys. J. B} {\bf 5}, 367 (1998).

\bibitem{Presura} C. Presura, D. van der Marel, A. Damascelli and
R.K. Kremer, {\it Phys. Rev.} {\bf B 61}, 15762 (2000).


\bibitem{Grenier}
B. Grenier, O. Cepas, L. P. Regnault, J. E. Lorenzo, T. Ziman,
J. P. Boucher, A. Hiess, T. Chatterji, J, Jegoudez and A. Revcolevschi,
{\it Phys. Rev. Letters} {\bf 86}, 5966 (2001). 



\end{thebibliography}
\end{document}